\documentclass[
  sort&compress,
  5p,
  preprint
]{elsarticle}

\usepackage{amsmath}            
\usepackage{braket}             
\usepackage{graphicx}           
\usepackage[utf8]{inputenc}     
\usepackage[draft]{hyperref}    
\usepackage{dcolumn}            
\usepackage{color}              
\usepackage{csquotes}           

\newcommand{\e}[1]{\text{e}^{#1}}
\newcommand{\vB}{\vec{B}}
\newcommand{\toG}{\Braket{\Omega | T\left\{\chi\left(\vr,t\right)\,\bar{\chi}\left(0\right)\right\}|\Omega}}
\newcommand{\pvB}{\psi_{\vB}}

\newcommand{\abs}[1]{\left|#1\right|}
\newcommand{\order}[1]{\mathcal{O}\left(#1\right)}
\newcommand{\aqeb}{\abs{qe\,B}}
\newcommand{\ab}{\abs{B}}
\newcommand{\epm}{\left(E\rb{B}+\mpi\right)}
\newcommand{\emm}{\left(E\rb{B}-\mpi\right)}

\newcommand{\vx}{\vec{x}}
\newcommand{\hmu}{\hat{\mu}}
\newcommand{\pivB}{\psi_{i,\vB}}
\newcommand{\ctdof}{\chi^2_{dof}}

\newcommand{\rb}[1]{\left(#1\right)}

\newcommand{\mpi}{m_\pi}

\newcommand{\pip}{\pi^+}
\newcommand{\api}{\alpha_{\pi}}
\newcommand{\bpi}{\beta_{\pi}}
\newcommand{\bpiz}{\beta^{\piz}}
\newcommand{\bpizu}{\beta^{\pizu}}
\newcommand{\bpizd}{\beta^{\pizd}}

\newcommand{\uu}{\overline{u}u}
\newcommand{\dd}{\overline{d}d}

\newcommand{\piz}{\pi^{0}}
\newcommand{\pizd}{\pi^{0}_{d}}
\newcommand{\pizu}{\piz_{u}}

\newcommand{\vr}{\vec{r}}

\newcommand{\eqnrtwo}[2]{Eqs.~$\left(\ref{#1}\right)$ and $\left(\ref{#2}\right)$}
\newcommand{\Fig}[1]{Figure \ref{#1}}
\newcommand{\Tab}[1]{Table \ref{#1}}

\newcommand{\Figtwo}[2]{Figures \ref{#1} and \ref{#2}}
\newcommand{\Refl}[1]{Ref.~\cite{#1}}    
\newcommand{\eqnr}[1]{Eq.~$\left(\ref{#1}\right)$}

\begin{document}

\title{Pion magnetic polarisability using the background field method}
\author{Ryan Bignell}
\ead{ryan.bignell@adelaide.edu.au}
\author{Waseem Kamleh}\ead{waseem.kamleh@adelaide.edu.au}
\author{Derek Leinweber}\ead{derek.leinweber@adelaide.edu.au}
\address{Special Research Centre for the Subatomic Structure of Matter (CSSM),\\
Department of Physics, University of Adelaide, Adelaide, South Australia 5005, Australia
}

\date{\today}

\begin{abstract}
  The magnetic polarisability is a fundamental property of hadrons, which provides insight into their
  structure in the low-energy regime. The pion magnetic polarisability is calculated
  using lattice QCD in the presence of background magnetic fields. The results presented are facilitated
  by the introduction of a new magnetic-field dependent quark-propagator eigenmode projector and the use of the background-field corrected clover fermion action. The magnetic polarisabilities are calculated in a relativistic formalism, and the excellent signal-to-noise property of pion correlation functions facilitates precise values.
\end{abstract}

\maketitle

\section{Introduction}
The electromagnetic polarisabilities of hadrons are of fundamental importance in the low-energy regime of quantum chromodynamics where they provide novel insight into the response of hadron structure to a magnetic field. The pion electric $\rb{\api}$ and magnetic $\rb{\bpi}$ polarisabilities are experimentally measured using Compton scattering experiments, such as $\gamma\,\pi \rightarrow \gamma\,\pi$~\cite{Antipov:1982kz,Adolph:2014kgj,Filkov:2018cey,Moinester:2019sew} where they enter into the description of the scattering angular distribution~\cite{Holstein:1990qy,Moinester:1997um,Scherer:1999yw,Ahrens:2004mg}.
\par
Theoretical approaches to calculating the pion electromagnetic polarisabilities are diverse. Calculations in the framework of chiral perturbation theory have a long history~\cite{Burgi:1996qi,Gasser:2006qa} while other approaches include dispersion sum rules~\cite{Filkov:1998rwz,Filkov:2008uyj,GarciaMartin:2010cw} and the linear $\sigma$ model~\cite{Bernard:1988gp}. Here we use the \emph{ab initio} formalism of lattice QCD with an external background field. This method involves direct calculation of pion energies in an external magnetic field where the relativistic energy-field relation~\cite{Martinelli:1982cb, PhysRevD.100.114518}
\begin{align}
  E^2\rb{B} = \mpi^2 &+ \rb{2\,n+1}\,\aqeb \nonumber \\
  &- 4\,\pi\,\mpi\,\bpi\,\ab^2 + \order{B^3},
  \label{eqn:E2B}  
\end{align}
can be used to extract the magnetic polarisability, $\bpi$. Here the pion has mass $\mpi$, charge $qe$ and the term proportional to $\aqeb$ is the Landau-level energy term~\cite{QFTZuber}. In principal there is an infinite tower of energy levels for $n=0,1,2,\dots$ but the lowest lying Landau level is isolated through Euclidean time evolution and a hadronic Landau-level projection for the charged pion. The background-field method has previously been used to extract the polarisabilities of baryons~\cite{Primer:2013pva, Bignell:2018acn} and nuclei~\cite{Chang:2015qxa} in dynamical QCD simulations as well as the magnetic polarisability of light mesons in quenched $SU(3)$ simulations~\cite{Lee:2005dq,Luschevskaya:2014lga,Luschevskaya:2015cko}. More recently the neutral $\pi$- and $\rho$-meson magnetic polarisabilities have been calculated in dynamical QCD~\cite{Bali:2015vua,Bali:2017ian,PhysRevD.100.114518,Ding:2020hxw}. In \Refl{Bali:2017ian}, Bali \textit{et al.} identified the spurious Wilson-fermion artefact associated with the background field. In \Refl{PhysRevD.100.114518}, using Wilson-clover fermions we introduced the Background-Field-Corrected clover fermion action which removes this spurious artefact.
\par
Background electric fields have also been used to calculate the electric polarisabilities of neutral hadrons such as the neutron~\cite{Freeman:2014kka} and neutral pion~\cite{Alexandru:2015dva}. Generalised background electromagnetic fields~\cite{Davoudi:2015cba} can be used to calculate diverse quantities such as nucleon spin polarisabilities~\cite{PhysRevD.47.3757,PhysRevD.73.114505} and the hadronic vacuum polarisation function~\cite{PhysRevD.92.054506}.
\par
The calculations presented herein are performed at several non-zero pion masses in order to motivate a chiral extrapolation to the physical regime. These polarisability values are provided with the intent of spurring future chiral effective field theory development to enable extrapolations to the physical regime incorporating finite-volume and sea-quark corrections.
\section{Simulation details \& background field method}
Four values of the light quark hopping parameter $\kappa_{ud}$ are used on the $2+1$ flavour dynamical gauge configurations provided by the PACS-CS~\cite{Aoki:2008sm} collaboration through the IDLG~\cite{Beckett:2009cb}. These provide pion masses of $\mpi = 0.702$, $0.572$, $0.411$ and  $0.296$ GeV. The lattice spacing varies slightly at each mass due to our use of the Sommer scale~\cite{Sommer:1993ce} with $r_0 = 0.49$. The lattice volume is $L^3 \times T = 32^3 \times 64$.
\par
The Background-Field-Corrected clover fermion action of \Refl{PhysRevD.100.114518} is used to remove spurious lattice artefacts that are introduced by the Wilson term . This action has a non-perturbatively improved clover coefficient for the QCD portion of the clover term and a tree-level coefficient for the portion deriving from the background field. This combination is effective in removing the additive-mass renormalisation induced by the Wilson term~\cite{Bali:2017ian}.
\par
To suppress wrap-around thermal effects, fixed boundary conditions are used in the temporal direction. We place the source at $N_t/4 =16$ and analyse correlation functions for $t \le 3\,N_t/4 = 48$ to ensure boundary effects in our correlation functions are negligible~\cite{Mahbub:2009nr}.
\par
No background field is present on the gaugefield ensembles and therefore this simulation is electroquenched. This is a departure from the physical world and should be accounted for in future chiral effective field theory work~\cite{Hall:2013dva}.
\subsection{Background field method}
The background field method~\cite{Smit:1986fn,Burkardt:1996vb,Davoudi:2015cba} induces a constant magnetic field by adding a minimal electromagnetic coupling to the (continuum) covariant derivative
\begin{align}
  D_{\mu} = D_{\mu}^{\text{QCD}} + i\,qe\,A_{\mu}.
\end{align}
This corresponds to a multiplication of the usual lattice QCD gauge links by an exponential phase factor
\begin{align}
  U_\mu(x) \rightarrow U^{\rb{B}}_\mu(x) = \e{i\,a\,qe\,A_\mu(x)}\,U_\mu(x),
  \label{eqn:UmuB}
\end{align}
where $a$ is the lattice spacing. For a uniform field along the $\hat{z}$ axis the spatial periodic boundary conditions induce a quantisation condition, limiting the choice of uniform magnetic field strengths to
\begin{align}
  \aqeb = \frac{2\,\pi\,k}{N_x\,N_y\,a^2},
  \label{eqn:qc}
\end{align}
where $k$ is an integer which governs the field strength for a particle of charge $qe$ and $N_x=N_y=32$ are lattice dimensions. The down quark has the smallest charge magnitude and governs the magnetic field quanta. As the $d$ quark has charge $q_{d}e$ the $\pip$ will have charge $q_{\pip}e = -3 \times q_{d}e$. That is, the smallest field strength for the $\pip$ has $k_{\pip}=-3$ and $k_d=1$.
\subsection{Quark Operators}
In this work a tuned Gaussian smeared source is used to provide a representation of QCD interactions.
The smearing level is varied at zero external field strength $(B=0)$ and the effective mass examined to determine the smearing which produces the earliest onset of plateau behaviour~\cite{Mahbub:2009nr}.
The resulting smearing levels are $N_{sm} = 150,\,175,\,300,\,250$ sweeps for ensembles with masses $\mpi = 0.702$, $0.572$, $0.411$, $0.296$ GeV respectively~\cite{Bignell:2020xkf}.
\par
As charged particles in an external magnetic field, the quarks will experience Landau type effects in addition to the confining force of QCD. To provide greater overlap with the energy eigenstates of the pion we use the $SU(3) \times U(1)$ eigenmode quark projection technique introduced in Ref.~\cite{Bignell:2020xkf}. In summary, the low-lying eigenmodes $\Ket{\psi_i}$ of the two-dimensional lattice Laplacian with both QCD and background field effects are calculated
\begin{align}
  \Delta_{\vx,\vx^\prime} = 4\,\delta_{\vx,\vx^\prime} -\! \sum_{\mu=1,2}U_\mu(\vx)\delta_{\vx+\hmu,\vx^\prime} + U^\dagger_\mu(\vx-\hmu)\delta_{\vx-\hmu,\vx^\prime}.
  \label{eqn:2DLap}
\end{align}
Here $U_\mu\rb{\vx}$ is the full $SU(3) \times U(1)$ gauge link of \eqnr{eqn:UmuB}.
\par
A projection operator can be defined by truncating the completeness relation $\mathcal{I} = \sum_{i=1}^n\,\Ket{\psi_i}\Bra{\psi_i}$. This truncation filters out the high-frequency modes, an effect similar to (2D) smearing.
In the pure $U(1)$ case each quark would have a definitive set of degenerate eigenmodes associated with each Landau level, however the introduction of QCD interactions into the Laplacian causes the $U(1)$ modes associated with the different Landau levels to mix~\cite{Bruckmann:2017pft}. It is clear that in the case of a charged hadron, it is the hadronic level Landau modes that are respected and as such there is no longer a single definite Landau mode that describes the quark-level physics in the confining phase.
We choose $n=96$ eigenmodes to construct the quark-level projection operator, in accordance with our previous study \Refl{Bignell:2020xkf} where this number was found to be sufficiently large as to avoid introducing significant noise into the correlation function whilst also small enough to place a focus on the low-energy physics relevant to the isolation of the magnetic polarisability.
\par
As the lattice Laplacian used is two-dimensional, the low-lying eigenspace for each $\rb{z,t}$ slice on the lattice is calculated independently, allowing for the four-dimensional coordinate space representation of an eigenmode
\begin{align}
  \Braket{\vx, t| \pivB} = \pivB\rb{x,y|z,t},
\end{align}
to be interpreted as selecting the two dimensional coordinate space representation $\pivB\rb{x,y}$ from the eigenspace belonging to the corresponding $\rb{z,t}$ slice of the lattice. Hence the four-dimensional coordinate space representation of the projection operator is
\begin{align}
  P_n\left(\vx,t;\vx^\prime,t'\right) = \sum_{i=1}^{n}\,\braket{\vx,t\,|\,\pivB}\braket{\pivB\,|\,\vx^\prime,t'}\,\delta_{zz'}\,\delta_{tt'},
  \label{eqn:coordP_n}
\end{align}
where the Kronecker delta functions ensure that the projector acts trivially on the $\rb{z,t}$ coordinates.
\par
This projection operator is then applied at the sink to the quark propagator in a coordinate space representation,
\begin{align}
  S_n\left(\vx,t;\vec{0},0\right) = \sum_{\vx^\prime}\,P_n\left(\vx,t;\vx^\prime,t\right)\,S\left(\vx^\prime,t;\vec{0},0\right).
\end{align}
The use of the $SU(3) \times U(1)$ eigenmode quark projection technique has introduced both QCD and magnetic field physics into the quark sink. This, along with a tuned smeared source produces pion correlation functions at non-trivial field strengths that have a strong overlap with the ground state pion, which occupies the lowest lying hadronic Landau level (as detailed in the next section).

\subsubsection{$U(1)$ Hadronic Landau Projection}
As a charged particle, the pion experiences hadronic level Landau effects, such that the ground state will occupy the lowest Landau level associated with the hadronic charge. In the presence of an external magnetic field along the $\hat{z}$ axis; the energy eigenstates of the $\pip$ are no longer eigenstates of the $p_x$ and $p_y$ momentum components.
\par
In a finite volume lattice the hadronic Landau levels correspond to the eigenmodes of the two-dimensional lattice Laplacian in Eq.~\ref{eqn:2DLap} where only the $U(1)$ background field is present. As $\abs{k_{\pi}}=\abs{3\,k_d}$, there is a degenerate subspace of $\abs{3\,k_d}$ eigenmodes to consider at the lowest hadronic Landau level, where $k_d$ is the down quark field quanta. We optimise a single $U(1)$ eigenmode, $\pvB\rb{x,y},$ to project the $(x,\,y)$ dependence of the two-point correlation function onto the lowest Landau level
\begin{align}
  G\left(p_z, \vB,t\right) = \sum_{\vr}\,&\pvB\left(x,y\right)\,\e{-i\,p_z\,z}\nonumber \\
  &\times \toG,
  \label{eqn:HadProj}
\end{align}
where $\vr = \rb{x,\,y,\,z}$. The eigenmode $\pvB\rb{x,y}$ in \eqnr{eqn:HadProj} is chosen to optimise the overlap with the source $\rho\rb{x,y} = \delta_{x,0}\,\delta_{y,0}$ (assumed to be at the origin) through a rotation of the $U(1)$ eigenmode basis that maximises the value of $\Braket{\rho|\pvB}$. An optional phase can be applied so that $\pvB\rb{0,0}$ is purely real at the source point.
\par
The projection of \eqnr{eqn:HadProj} is critical to successful isolation of the $\pip$ energy-eigenstate in a background magnetic field~\cite{Tiburzi:2012ks}.
\section{Magnetic Polarisability}
Defining the following combinations of two-point correlation functions
\begin{align}
  R_{+}\rb{B,t} &= G\rb{B,t}\,G\rb{0,t}, \label{eqn:R+} \\
  R_{-}\rb{B,t} &= \frac{G\rb{B,t}}{G\rb{0,t}}, \label{eqn:R-}
\end{align}
where $G\rb{B,t}$ is the correlation function for $p_z = 0$ in a magnetic field of strength B, then the energy shift is simply
\begin{align}
  \epm&\,\emm = E^2\rb{B} - \mpi^2 \nonumber \\&= \aqeb - 4\,\pi\,\mpi\,\bpi\,\ab^2 + \order{B^3}.
  \label{eqn:Eshift}
\end{align}
Specifically, the effective energies
\begin{align}
  E(\vB,t) \pm \mpi = \frac{1}{\delta t}\,\text{log}\rb{ \frac{ R_{\pm}(\vB,t) }{ R_{\pm}(\vB,t+\delta t) }},
\end{align}
are calculated with $\delta t = 2$. This formulation advantageously removes a portion of correlated QCD fluctuations, allowing the magnetic polarisability to be extracted using a simple polynomial fit. In order to constrain the charge of the pion to be $q = 1$, the fit performed is
\begin{align}
  E^2\rb{k_d} - \mpi^2 - \abs{a\,B} = c_2\,k_d^2 + \order{B^3},
  \label{eqn:E2fit}
\end{align}
where $c_2$ has the units of $E^2\rb{k_d}$ and is the fit parameter which is related to the magnetic polarisability using \eqnrtwo{eqn:qc}{eqn:Eshift}
\begin{align}
  \beta = -c_2\,\alpha\,\frac{q_d^2\,a^2}{\mpi}\,\rb{\frac{N_x\,N_y}{2\,\pi}}^2,
\end{align}
where $\alpha = 1/137\dots$ is the fine structure constant.
\subsection{Fitting}
\begin{figure}[tb]
  \includegraphics[width=\columnwidth]{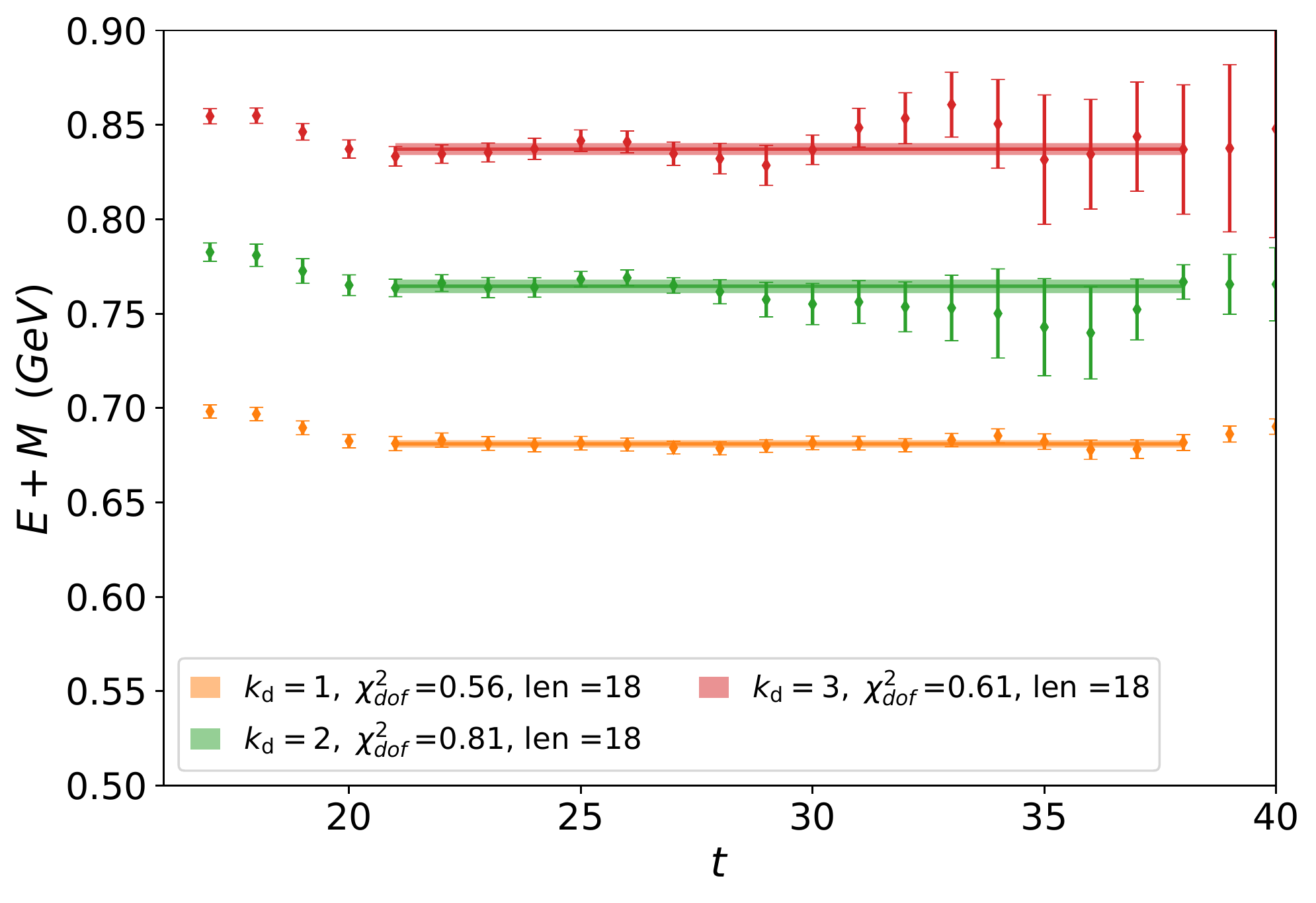}%
  \caption{\label{fig:EplusM:k13770}$\pip$ energy shift $\epm$ using \eqnr{eqn:R+} for the $\mpi = 0.296$ GeV ensemble.}
\end{figure}
\begin{figure}[tb]
  \includegraphics[width=\columnwidth]{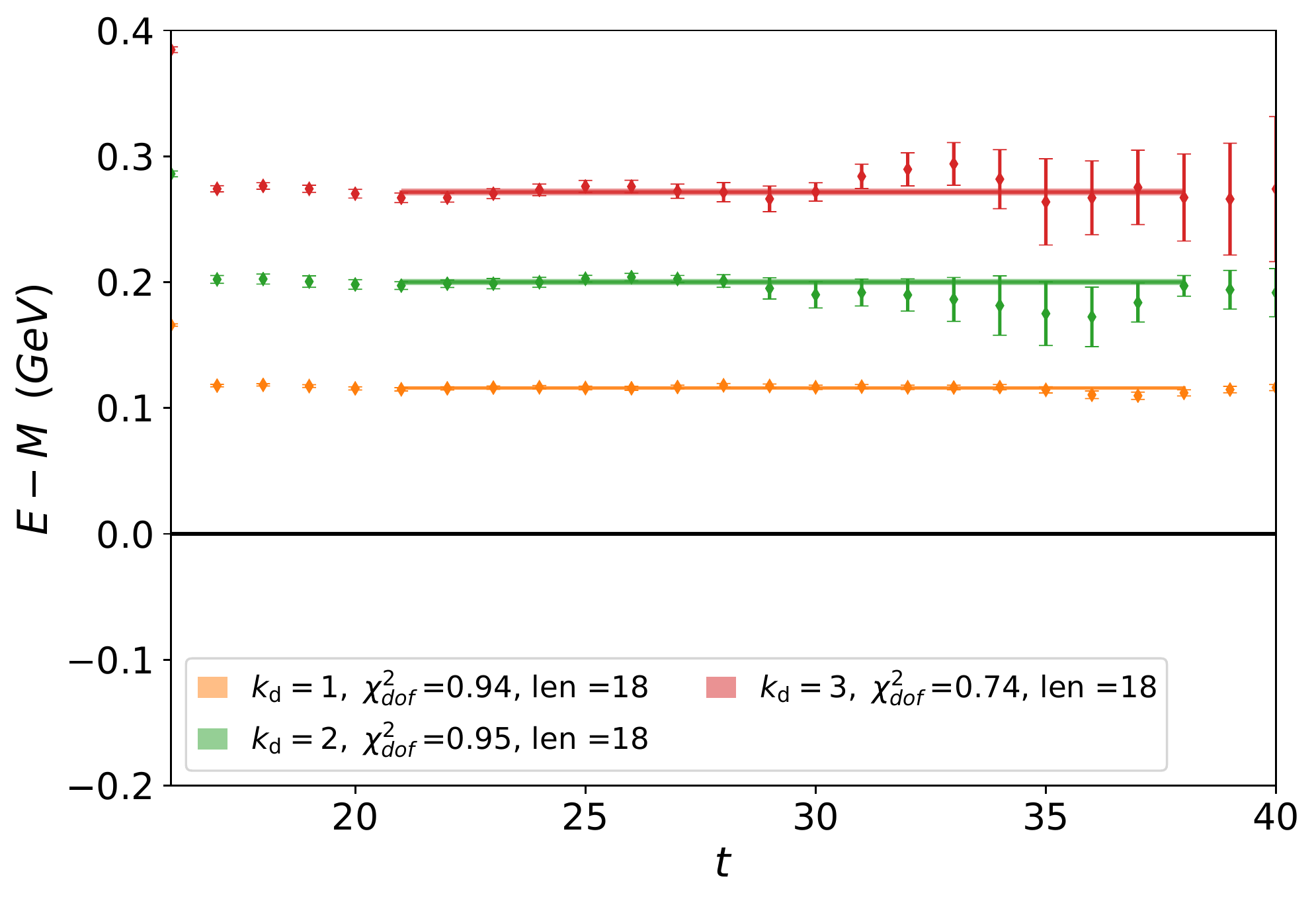}%
  \caption{\label{fig:deltaEpol:k13770}$\pip$ energy shift $\emm$ using \eqnr{eqn:R-} for the $\mpi = 0.296$ GeV ensemble.}
\end{figure}
The two effective-energy shifts $\epm$ and \\$\emm$ generated by the correlator combinations of \eqnrtwo{eqn:R+}{eqn:R-} are required to have plateau behaviour reflecting an isolated energy eigenstate. This isolation is evident in the long constant fits in \Figtwo{fig:EplusM:k13770}{fig:deltaEpol:k13770} for the $\mpi = 0.296$ GeV pion. The isolation is a result of our detailed projection treatment of the quark level effects of the background field. This is the first time that plateau behaviour has been observed in these quantities.
\begin{figure*}[t]
        \centering
        \includegraphics[width=0.475\textwidth]{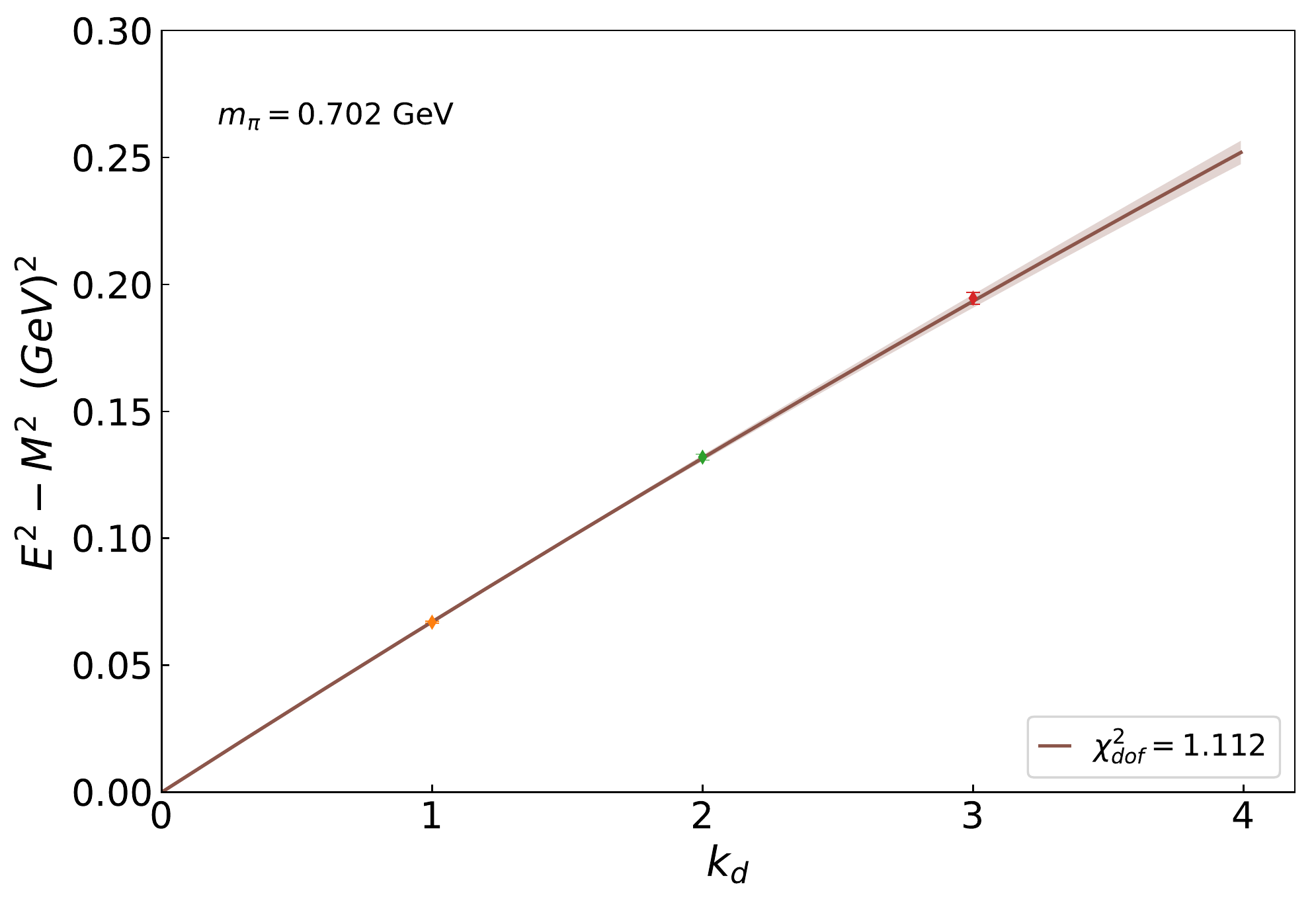}
	\includegraphics[width=0.475\textwidth]{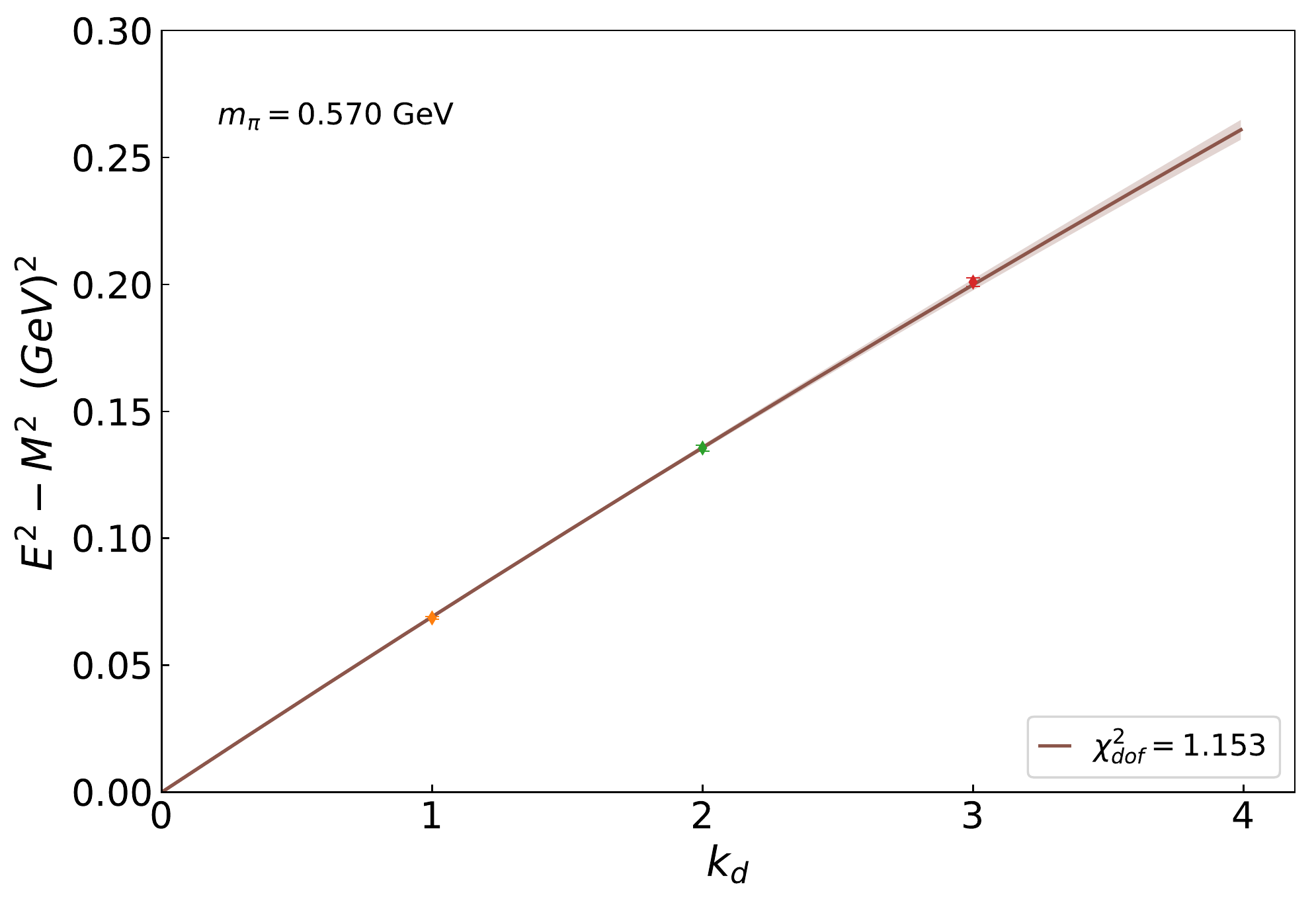}
        \\
        \includegraphics[width=0.475\textwidth]{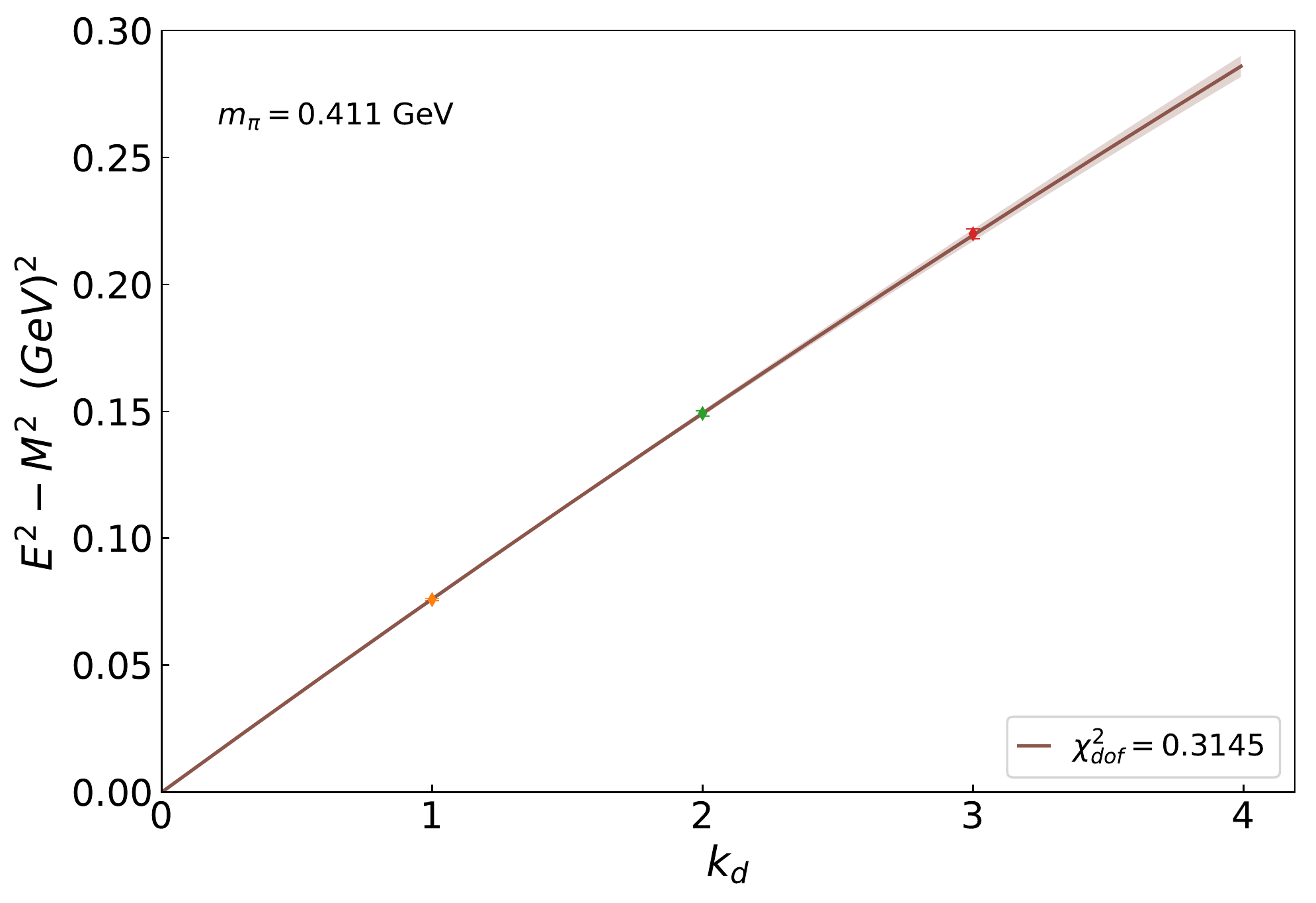}
        \includegraphics[width=0.475\textwidth]{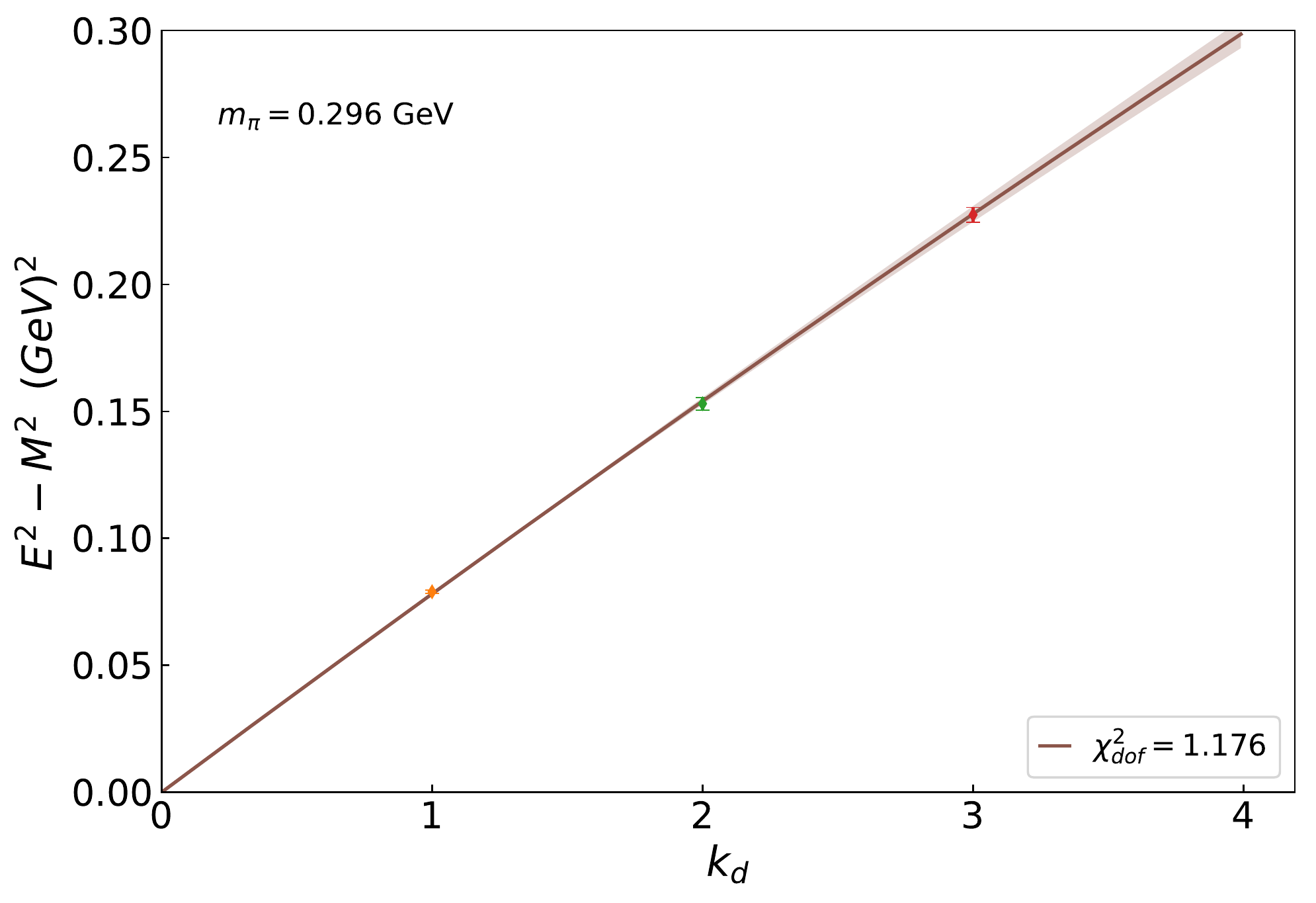}
        \caption{Linearly constrained quadratic fits of the energy shift of \eqnr{eqn:Eshift} to the field quanta at each quark mass for the $\pip$.}
	\label{fig:dEfits4-pip}
\end{figure*}
\par
From here the energy shift and fits of \eqnr{eqn:E2fit} are performed and the magnetic polarisability extracted. The fit function of \eqnr{eqn:E2fit} is considered, with the fits selected through a consideration of the full covariance matrix $\ctdof$. The selected fits are displayed in \Fig{fig:dEfits4-pip}, where the fit for $\kappa_{ud} = 0.13770$ of the $\mpi = 0.296$ GeV ensemble corresponds to the fit window displayed in \Figtwo{fig:EplusM:k13770}{fig:deltaEpol:k13770}. This is the first lattice calculation in which the fully relativistic energy shift of \eqnr{eqn:E2B} has been used. This is made possible due to the enhanced precision of pion correlation functions from the $SU(3) \times U(1)$ eigenmode projected quark propagator and Landau-projected hadron sink..
\par
The neutral pion is also amenable using these techniques. Here we consider the neutral, source-sink connected pion with quark content $\dd$. The fit of \eqnr{eqn:E2fit} now needs no explicit subtraction of the Landau energy term as the $\piz_{d}$ is overall charge-less. The success of the quadratic only and linearly constrained quadratic fits to the highly precise $\piz_{d}$ and $\pi^{+}$ energy shifts of \eqnr{eqn:Eshift} suggests that higher order contributions in $B$ are negligible. These neutral pion results draw from these new techniques, in particular the inclusion of the $SU(3) \times U(1)$ quark-propagator Laplacian projection which enables improved energy shift plateaus to be fitted.
\par
The magnetic polarisability of the $\piz$ may be estimated by considering the average of the magnetic polarisability of the $\uu$ and $\dd$ pions
\begin{align}
  \bpiz = \frac{1}{2}\,\rb{\bpizd + \bpizu},
\end{align}
where $\pizu$ is the pion with quark content $\uu$. This pion has relativistic energy
\begin{align}
  E^2_{\pizu}\rb{B} - \mpi^2 = -4\,\pi\,\mpi\,\bpizu\,\ab^2 + \order{B^3}.
\end{align}
As the $\uu$ pion is simply the $\dd$ pion in a field of twice the magnitude
\begin{align}
  E^2_{\pizu}\rb{\frac{B}{2}} = E^2_{\pizd}\rb{B},
  \label{eqn:udequiv}
\end{align}
we may write
\begin{align}
  E^2_{\pizu}\rb{\frac{B}{2}} - \mpi^2 &= -4\,\pi\,\mpi\,\bpizu\,\abs{\frac{B}{2}}^2 + \order{B^3} \nonumber \\
  &=-4\,\pi\,\mpi\,\bpizu\,\frac{1}{4}\,\abs{B}^2 + \order{B^3},
  \label{eqn:bpizu}
\end{align}
and hence
\begin{align}
  E^2_{\pizd}\rb{B} - \mpi^2 &= -4\,\pi\,\mpi\,\bpizd\,\abs{B}^2 + \order{B^3} \nonumber \\
  &=-4\,\pi\,\mpi\,\bpizu\,\frac{1}{4}\,\abs{B}^2 + \order{B^3},
\end{align}
where we have used \eqnrtwo{eqn:udequiv}{eqn:bpizu}. Thus the magnetic polarisability of the $\uu$ pion is related to that of the $\dd$ pion by
\begin{align}
  \bpizu &=4\,\bpizd.
\end{align}
The magnetic polarisability of the full neutral pion is then estimated as
\begin{align}
  \beta^{\piz} &= \frac{1}{2}\,\rb{\bpizd + \bpizu} = \frac{5}{2}\,\bpizd.
  \label{eqn:fullbetapiz}
\end{align}
Our resulting pion magnetic polarisabilities are presented in \Tab{tab:ResTab}.
\begin{table}[]
  \centering
  \caption{Magnetic polarisability values for the pion at each quark mass considered. The numbers in parantheses describe statistical uncertainties.}
  \resizebox{\columnwidth}{!}{%
    \begin{tabular}{cccccc}
      \hline\hline
      $\kappa_{ud}$ & $\mpi$ (GeV) & $a$ (fm) & $\beta^{\pip}$ ($\times 10^{-4}$ fm$^3$) & $\bpizd$ ($\times 10^{-4}$ fm$^3$) &$\bpiz$ ($\times 10^{-4}$ fm$^3$)\\ \hline
      0.13700       & 0.702          & 0.1023   & 0.255(56)                                & 0.900(17)                          & 2.25(5)      \\
      0.13727       & 0.570          & 0.1009   & 0.275(54)                                & 0.872(16)                          & 2.18(4)      \\
      0.13754       & 0.411          & 0.0961   & 0.355(62)                                & 0.766(33)                          & 1.92(9)      \\
      0.13770       & 0.296          & 0.0951   & 0.35(11)                                 & 0.754(35)                          & 1.89(9)      \\\hline\hline
    \end{tabular}%
  }
  \label{tab:ResTab}
  \end{table}
\begin{figure}
  \includegraphics[width=\columnwidth]{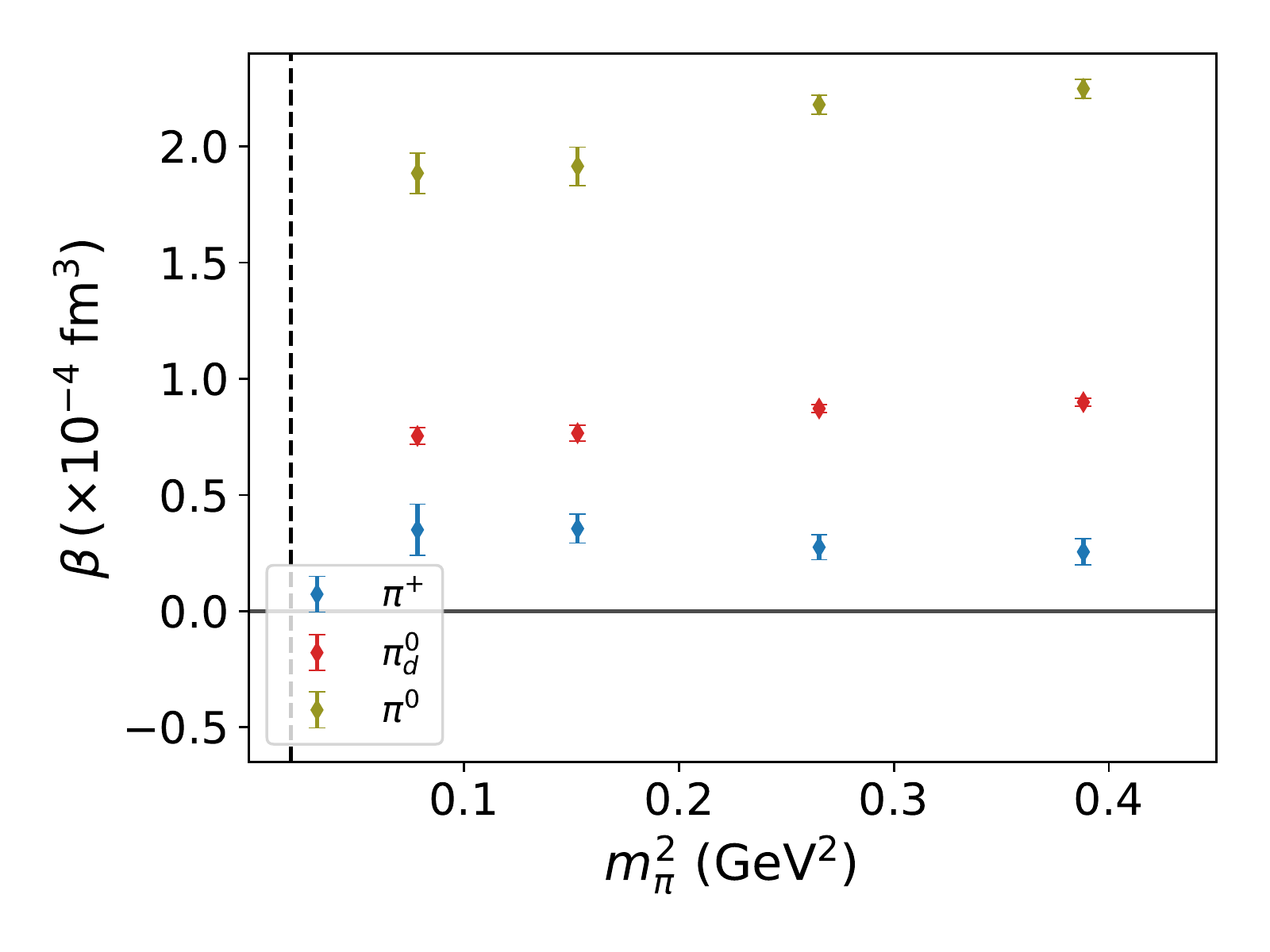}%
  \caption{\label{fig:betaMpi}The magnetic polarisability of the pion from lattice QCD is plotted as a function of $\mpi^2$.}
\end{figure}
\par
All quark masses produce similar values for $\beta^{\pip}$ and $\beta^{\piz}$. This is in contrast to the neutron~\cite{Bignell:2018acn} and evident in \Fig{fig:betaMpi} where our magnetic polarisabilities are plotted as a function of pion mass squared. The neutral pion results using \eqnr{eqn:fullbetapiz} are in good agreement with a number of theoretical approaches and experimental measurements~\cite{Gasser:2006qa,Filkov:2005ccw}. \Refl{Ding:2020hxw} presents results for $\beta^{\piz}$ and $\beta^{\pizd}$ in their Table II which have a ratio consistent with \eqnr{eqn:fullbetapiz}.
\par
We note again that we consider only the source-sink connected portion of the neutral pion correlator here. Pion electromagnetic polarisabilities were studied using chiral perturbation theory for partially quenched QCD in \Refl{Hu:2007ts} where the polarisability of the neutral pion at one-loop order arises entirely from self-annihilation contractions. These terms are also expected to scale quadratically with the field strength and so may provide an important correction in the full magnetic polarisability of the neutral pion.
\par
The results presented herein utilise the $SU(3) \times U(1)$ eigenmode quark-projection technique that we introduced in \Refl{Bignell:2020xkf}. The success of this technique is evident in the improved energy-shift plateaus when compared to the equivalent energy shifts of \Refl{PhysRevD.100.114518}. These new results represent an improved understanding of the physics relevant to the extraction of the magnetic polarisability of the pion.

\section{Conclusion}
The magnetic polarisability of the charged pion has been calculated using lattice QCD for the first time. This is an important step forward in our understanding of this fundamental property, made possible due to the use of the $SU(3) \times U(1)$ eigenmode projection technique, along with a hadronic Landau eigenmode projection. The neutral pion magnetic polarisability is also presented. These results represent the first systematic study of pion magnetic polarisabilities across a range of pion masses with a fermion action which does not suffer from magnetic-field dependent quark-mass renormalisation effects.
\par
To connect these results to experiment, one can draw on chiral effective field theory. By formulating the theory in a finite volume, finite-volume corrections can be determined. Moreover, by separating the contributions of valence and sea quarks, using the techniques of partially-quenched chiral effective field theory, one can address the electro-quenched aspects of this calculations. Thus our results present an interesting challenge for the effective field theory community.
\par
Future work in lattice QCD for the pion magnetic polarisabilities could focus on calculating the full neutral pion correlator which includes self-annihilation contractions and thus requires the $x$-to-$x$ loop propagator~\cite{Foley:2005ac}. Similarly the electroquenched nature of our calculations could be addressed by extending the background field to the \enquote{sea} quarks of the simulation at gaugefield generation time. Such a calculation requires a separate set of gaugefields for each external magnetic field strength and thus removes the advantageous QCD correlations between two-point correlation functions at zero and finite external field strengths. Reweighting~\cite{Freeman:2014kka} promises an avenue to preserve these correlations.

\section*{Acknowledgements}
We thank the PACS-CS Collaboration for making their $2+1$ flavour configurations available and the ongoing support of the International Lattice Data Grid (ILDG). We also thank Heng-Tong Ding for interesting exchanges on the neutral pion. This work was supported with supercomputing resources provided by the Phoenix HPC service at the University of Adelaide. This research was undertaken with the assistance of resources from the National Computational Infrastructure (NCI). NCI resources were provided through the National Computational Merit Allocation Scheme, supported by the Australian Government through Grants No. LE190100021, LE160100051 and the University of Adelaide Partner Share.  R.B. was supported by an Australian Government Research Training Program Scholarship.  This research is supported by the Australian Research Council through Grants No. DP140103067, \\DP150103164, DP190102215 (D.B.L) and DP190100297 (W.K).
%


%
\bibliography{pip.bib}
\bibliographystyle{elsarticle-num-names}
\end{document}